\documentclass[aps,superscript address,float fix,two column,reprint,footinbib]{revtex4-1} 

\usepackage{amsmath,amssymb,bm,epsfig}
\usepackage{natbib}
\usepackage{hyperref} 
\usepackage{graphicx}
\usepackage{wasysym}
\RequirePackage{lineno}

\usepackage[utf8]{inputenc}

\usepackage{xcolor,colortbl}
\usepackage{pifont}
\usepackage{footnote}

\def\ptmin{p^{\textrm{min}}_T}

\def\Nch{N_{\rm ch}^{\rm rec}}

\def\Ntrig{N_{\rm trig}}

\def\as{\alpha_S}
\def\sp{S_\perp}
\def\qs{Q_{\!_S}}
\def\qnot{Q_{\!0}}
\def\Qs{\qs}
\def\nc{N_c}
\def\cf{C_F}
 
\def\qp{ {\bf q}_{_T} } 
 
\def\pp{ {\bf p}_{_T} } 
\def\kp{ {\bf k}_{_T} } 
 
\newcommand{\kpn}[1]{ {\bf k}_{#1\perp} } 

\newcommand{\ud}{\mathrm{d}}

\def \beq{\begin{equation}}
\def \eeq{\end{equation}}
\def \beqa{\begin{eqnarray}}
\def \eeqa{\end{eqnarray}}

\begin{document}

\title{Energy dependence of the ridge in high multiplicity proton-proton collisions}

\author{Kevin Dusling}
\affiliation{American Physical Society, 1 Research Road, Ridge, NY 11961, USA}
\affiliation{Physics Department, Brookhaven National Laboratory,
  Upton, NY 11973, USA
}

\author{Prithwish Tribedy}
\affiliation{Physics Department, Brookhaven National Laboratory,
  Upton, NY 11973, USA
}

\author{Raju Venugopalan}
\affiliation{Physics Department, Brookhaven National Laboratory,
  Upton, NY 11973, USA
}
\affiliation{Institut f\"{u}r Theoretische Physik, Universit\"{a}t Heidelberg, Philosophenweg 16, 69120 Heidelberg, Germany
}

\begin{abstract}
We demonstrate that the recent measurement of azimuthally collimated, long-range rapidity (``ridge") correlations in $\sqrt{s}=13$ TeV proton-proton (p+p) collisions by the ATLAS collaboration at the LHC are in agreement with expectations from the color glass condensate effective theory
of high energy QCD.  The observation that the integrated near side yield as a function of multiplicity is
independent of collision energy  is a natural consequence of the fact that multiparticle production is driven by a single semihard saturation scale in the color glass condensate framework.  We argue further  that the azimuthal structure of these recent ATLAS ridge measurements strongly constrain hydrodynamic interpretations of such correlations in high multiplicity p+p collisions. 
\end{abstract}
\maketitle

The systematics of azimuthally correlated, long-range rapidity correlations--the so-called ridge correlations--have recently been measured in small systems ranging from proton-proton (p+p), proton-nucleus (p+A), deuteron-nucleus (d+A) and Helium 3-nucleus ($^3$He+A) collisions at both RHIC and LHC~\cite{Khachatryan:2010gv, CMS:2012qk, Abelev:2012ola, Aad:2012gla, Adare:2014keg, Adamczyk:2015xjc, Adare:2015ctn}. An outstanding question is whether such correlations arise from the final state hydrodynamic flow of the produced matter--as is believed to be the case for nucleus-nucleus (A+A) collisions--or from correlations that are already embedded in the initial state wavefunctions of the projectile and target. 

In a series of papers, two of us argued that many features of the ridge correlations in high multiplicity p+p collisions could be quantitatively explained within the initial state framework of the color glass condensate (CGC) effective theory of high energy
QCD~\cite{Dusling:2012iga,Dusling:2012cg,Dusling:2012wy,Dusling:2013qoz}. Our results were derived within a ``glasma graph" approximation to the most general set of possible two particle correlation Feynman diagrams that contribute within the power counting scheme established by the CGC effective theory. 

Subsequent to the early p+p high multiplicity ridge measurements, ridge correlations observed in light-heavy ion collisions at both the RHIC and LHC appear to indicate a collective origin of some of the systematics exhibited by the measurements~\cite{Bozek:2013uha,Bozek:2013ska,Bzdak:2013zma,Qin:2013bha,Schenke:2014zha,Bzdak:2014dia,Werner:2013ipa}.  However some of these systematics adduced to support the hydrodynamic flow interpretation may also arise within initial state frameworks~\cite{Kovner:2010xk,Kovner:2011pe,Kovchegov:2012nd,Kovchegov:2013ewa,Dumitru:2014dra,Dumitru:2014vka,Dumitru:2014yza,Gyulassy:2014cfa,
Lappi:2015vha,Schenke:2015aqa} and others, notably the hadron fragmentation patterns often attributed to hydrodynamical flow, may have alternative interpretations~\cite{Zhou:2015iba,Ortiz:2013yxa}. We shall not discuss these further here but merely note that the interpretations of the origins of the ridge correlations remain fluid;   further data is urgently needed to help sort through the competing (and compelling) explanations.

Recently the ATLAS Collaboration reported first measurements of ridge correlations in high multiplicity p+p collisions at the LHC with the highest center-of-mass energy of 13 TeV. They found quite strikingly that the associated integrated yield of the ridge collimation agrees (within experimental errors) with the CMS results at the lower 7~TeV center-of-mass energy when plotted as a function of multiplicity~\cite{1384640}.
We will demonstrate here that the observed energy independence of the ridge is a natural consequence of the fact that multiparticle production in the color glass condensate is controlled by a single semihard saturation scale $Q_S$ in collisions of symmetric systems like p+p. More concretely, we will show that the ATLAS results are in quantitative agreement with our prior computations with no adjustment of parameters~\footnote{All the parameters in this study are identical to those
used in~\cite{Dusling:2013qoz} for p+p collisions at $\sqrt{s}$= 7~TeV. The
only difference in this work relative to our previous study is an improved
extrapolation of the unintegrated gluon distributions at large~$x$ and $k_\perp$. This improvement has a negligible effect on the observables studied here and in prior works.}.

Before we proceed to discuss the details of the glasma graph computation, it is useful to keep in mind a few generic features of multiparticle production in this framework. 
The azimuthal structure of the two particle correlations results from an interplay between mini-jet (more generically di-jet) production along with glasma graph contributions at high gluon densities~\cite{Dusling:2012cg,Dusling:2012wy}. 

Di-jet production proceeds via single gluon exchange in the $t$-channel and is the dominant process for di-hadron production in minimum bias events corresponding to low parton densities.  As the two gluons are produced from the same $t$-channel gluon exchange, the correlation is kinematically constrained to be back-to-back in their relative azimuthal angle ($\Delta\Phi\approx \pi$). In contrast, the glasma graphs contain two independent ladders and are therefore not similarly constrained; they are however highly suppressed for high transverse momenta and low parton densities.  It was argued however~\cite{Dumitru:2008wn,Dumitru:2010iy}, based on the power counting of the CGC effective theory, that glasma graphs can be enhanced by $\as^{-4}$ with respect to the di-jet contributions at high parton densities. The fact that the proton unintegrated gluon distributions are peaked about transverse momenta $Q_S$ in the CGC effective theory, and the kinematical constraints that the two gluons are produced with momenta $p_T\sim q_T \sim Q_S$ generates angular correlations that are symmetric around $\Delta\Phi=\pi/2$ and enhanced at $\Delta\Phi \approx 0$ and $\Delta\Phi\approx \pi$. The glasma graph two particle correlations therefore provide a natural source of the symmetric near side and away side ridge in high multiplicity collisions.

We note further that, because of the azimuthal symmetry noted, only positive even Fourier coefficients~\footnote{That the Fourier coefficients are even is manifest in the glasma framework. That they are positive is not but is true for the kinematics explored in this work.} of the azimuthal dependence of the two particle glasma correlation can be entertained in the glasma graph approximation. On the other hand, the Fourier decomposition of the di-jet gives both even and odd Fourier coefficients. The restriction of the glasma correlation to even Fourier coefficients is a consequence of the glasma graph approximation which is valid for $Q_S^2\gg \Lambda_{\rm QCD}^2$ but $Q_S \lesssim p_T,q_T$, where $\Lambda_{\rm QCD}\sim 100-200$ MeV is the intrinsic QCD scale. For $p_T,q_T \lesssim Q_S$, positive odd moments in the Fourier decomposition of the two gluon azimuthal correlation are also feasible~\cite{Lappi:2009xa,Schenke:2015aqa} due to multiple scattering contributions that break the charge and parity invariance manifest in the glasma graph computation. Thus depending on the values of $Q_S$ and $p_T,q_T$ examined, an interesting pattern of positive and negative Fourier modes of the azimuthal correlations may be anticipated. In contrast, hydrodynamics will generically give Fourier coefficients of the two particle correlation that are positive definite~\footnote{These Fourier coefficients must be distinguished from the Fourier coefficients of the single particle azimuthal anisotropy measured with respect to a reaction plane}. We will return to this point later on in the manuscript.

Last but not least, the glasma framework is straightforwardly extended to three particle
correlations~\cite{Dusling:2009ar} and generalized to $n$-particle correlations as well~\cite{Gelis:2009wh}. In the latter case, the resulting cumulants naturally 
result in a multiplicity distribution having the form of a negative binomial
distribution.  The tail of this distribution corresponds to the rare high
multiplicity events relevant for our discussion~\footnote{This picture of multiparticle production is an 
essential ingredient in the implementation of realistic initial conditions for the event-by-event hydrodynamics of A+A collisions, as developed in particular in the IP-Glasma model~\cite{Schenke:2012wb, Schenke:2012fw}.}. This point too shall be revisited before we conclude our discussion. 

Now turning to the glasma graph computation, we first note that the double inclusive gluon distribution is approximated by a $k_T$-factorized form~\footnote{For a comprehensive discussion of various approximation schemes in the CGC effective theory, and their comparison to numerical simulations, we refer the reader to \cite{Lappi:2015aa}.}.  There are a total of eight glasma graphs that are used in this analysis; the full expressions can be found in~\cite{Dusling:2013qoz}. Let us give as an example the expression from one such diagram responsible for the near side ridge
\begin{align}
\label{eq:glasma}
\frac{d^2N_{\rm \sl Glasma1}^{\rm \sl corr.}}{d^2\pp d^2\qp dy_p dy_q}&=
\frac{32\as^2}{(2\pi)^{10}\zeta\;\nc\cf^3}\,\frac{S_\perp}{\pp^2\qp^2}\times\\
& \int \limits_{\kp} \Phi_A^2(\kp)\Phi_B(\pp-\kp)\Phi_B(\qp-\kp)\;,\nonumber 
\end{align}
where the unintegrated gluon distributions (UGDs) $\Phi$ are evaluated at (small) $x$ values on the order $x\sim p_T/\sqrt{s}e^{\pm y_p}$ -- the precise prescription is given in the reference above.  The unintegrated gluon distributions can be expressed in terms of ${\cal T}(x,r_\perp)$--the forward scattering amplitude of a quark-antiquark dipole of transverse size $r_\perp$ on a proton/nuclear target-through the expression
\begin{equation}
\Phi(x, k_\perp) = {\pi \nc k_\perp^2\over 2\as}\int \limits_0^\infty dr_\perp r_\perp J_0(k_\perp r_\perp)  [1-{\cal T} (x,r_\perp)]^2.
\label{eq:ugd}
\end{equation}
The unintegrated gluon distribution at a given momentum fraction $x$ is obtained by solving the Balitsky-Kovchegov (BK) renormalization equation~\cite{Balitsky:1995ub, Kovchegov:1999yj}. The implementation we use for our study includes the running coupling next-to-leading-log (NLL) correction to BK, hereafter referred as the rcBK equation~\cite{Balitsky:2006wa}. The initial condition for ${\cal T}_{A,B} (x,r_\perp)$ in the rcBK equation is set according to the McLerran-Venugopalan (MV) model~\cite{McLerran:1993ni,McLerran:1993ka} with a finite anomalous dimension. In the MV model, the shape of ${\cal T}_{A,B} (x,r_\perp)$ is expressed as a function of the dimensionless quantity $r_\perp^2 \qnot^2$, where $\qnot^2$ is a nonperturbative scale. Fixing the value of $\qnot^2$ prescribes the initial shape of ${\cal T}_{A,B} (x,r_\perp)$ at an initial $x\!=\!x_0$. However such an initialization is not unique since the scattering amplitude is known to have a strong dependence on the impact parameter~\cite{Kowalski:2006hc}. Such a dependence can be accounted for by varying the value of $\qnot^2$ to initialize ${\cal T}_{A,B} (x,r_\perp)$. 

The value of $\qnot^2=0.168$ GeV$^2$ (in the fundamental representation) describes the minimum bias~\cite{Kowalski:2006hc,Rezaeian:2012ji} HERA deeply inelastic electron-proton scattering data. Therefore higher $\qnot^2$ will correspond to the scattering of a color dipole off the proton at smaller impact parameters where the color charge densities in the target are higher. Even if a fixed impact parameter is probed by the dipole, there could be fluctuations in the configurations of the color charge inside the proton purely driven by stochastic processes that are nonperturbative at the initial scale~\cite{Iancu:2004es,Iancu:2004iy}. 

The effect of such fluctuations will give rise to a distribution of $\qnot$'s. By choosing higher $\qnot^2$'s, one can therefore incorporate the physics of rare configurations of color charge densities~\cite{Dusling:2013qoz,McLerran:2015qxa}. We will therefore model rare high multiplicity p+p collisions by using UGDs corresponding to higher values of $\qnot^2$. It is important to note here that the  expressions~Eq.\ref{eq:glasma} and Eq.\ref{eq:bfkl} are inclusive quantities and, strictly speaking, high multiplicity events do not necessarily correspond to using a larger initial $\qnot$ in the inclusive yield.
In a fully first principles approach, in order to get the
multiplicity distribution right, one must include impact parameter
fluctuations, color charge fluctuations (which generate a negative-binomial
distribution) and intrinsic gluon fluctuations of the initial saturation
momentum (See \cite{McLerran:2015qxa} for more details). Such a first
principles treatment, combined with small-x evolution (here
included via running coupling BK), is currently out of reach in high energy
QCD.  We therefore resort to adjusting the initial saturation scale $\qnot$ to
the match the multiplicity of interest. 
 
However once this initial condition is set, the quantum evolution in $x$ is uniquely given by the rcBK equation which evolves ${\cal T}_{A,B} (x,r_\perp)$ (and therefore the UGDs) to smaller values of $x$. The peak of the distribution $\Phi_{A,B} (x, k_\perp)$ corresponds to the saturation scale $\qs^2$ at that $x$.  For the typical kinematics probed by these experiments, we find saturation scales (in the adjoint representation) on the order of $\Qs\sim 1.5~(3.0)$~GeV for min-bias (central) collisions.  It must be noted that the UGDs that go into the calculation of the two particle correlation do not include any free parameters and are well constrained by HERA data. 

In addition to the just discussed UGDs, Eq.~\ref{eq:glasma} depends on the transverse overlap area $S_\perp$ and the nonperturbative constant $\zeta$. The latter estimates contributions to multiparticle production below the scale $\qs^2$. The transverse overlap area $S_\perp$ cancels out in the expression for the per-trigger yield.  This leaves only the parameter $\zeta$, which was extracted previously by an independent $\kp$ factorization analysis of data on $n$-particle multiplicity distributions~\cite{Tribedy:2011aa,Tribedy:2010ab} and estimated to be $\zeta=1/6$. Numerical simulations by solving classical Yang-Mills also constrain the value of $\zeta$ to be of similar order~\cite{Lappi:2009xa, Schenke:2012fw}. We should emphasize that precisely this value was used in our prior fits to the p+p CMS 7 TeV data. 

The di-jet contribution in this framework is estimated ~\cite{Colferai:2010wu, Fadin:1996zv} to be 
\begin{eqnarray}
\label{eq:bfkl}
&&\frac{d^2N_{\rm \sl BFKL}^{\rm \sl corr.}}{d^2\pp d^2\qp dy_p dy_q} = \frac{32\,\nc\,
\alpha_s^2}{ (2\pi)^8 \,\cf}\,\frac{\sp}{\pp^2\qp^2}\times \\
&&\int \limits_{{\kpn{0}}} \! \int \limits_{{\kpn{3}}} \!\!
\Phi_A(\kpn{0})\Phi_B(\kpn{3})\,\mathcal{G}(\kpn{0}\!\!-\!\!\pp,\kpn{3}\!\!+\!\!\qp,y_p\!\!-\!\!y_q) ,\nonumber
\end{eqnarray}
where $\cf=(\nc^2-1)/2 \nc$ and $\mathcal{G}$ is the BFKL Green's function that generates gluon emissions between the gluons that fragment into triggered hadrons. The detailed NLL  expression of $\mathcal{G}$ can be found in Ref~\cite{Dusling:2013qoz}. As previously, the transverse overlap area $S_\perp$ in Eq.~\ref{eq:bfkl} cancels in the final expression of per-trigger di-hadron yield. 

The combined effect of the glasma graph and di-jet contributions to the total two particle correlation is given by 
\begin{align}
\label{eq_twocomp}
\frac{d^2N^{\rm \sl corr.}}{d^2\pp d^2\qp dy_p dy_q} =\hspace{.5cm}& \\
\frac{d^2N_{\rm \sl Glasma}^{\rm \sl corr.}}{d^2\pp d^2\qp dy_p dy_q} &+ \frac{d^2N_{\rm \sl BFKL}^{\rm \sl corr.}}{d^2\pp d^2\qp dy_p dy_q}.\nonumber
\end{align}
The decomposition of the above double-inclusive distribution into a sum of glasma and BFKL (di-jet) contributions is far from trivial. The di-jet diagram (BFKL) is a one-loop correction (and therefore $\as$ suppressed but
enhanced by $\nc^2$) with respect to the Glasma diagrams.  Its full calculation
in the ``dense-dense” limit of high multiplicity collisions requires knowledge
of the background-field gluon propagator in these collisions, which is not yet
available.  

Nevertheless, significant work has been done in the dilute dense limit of the CGC \cite{Baier:2005dv,Fukushima:2008ya}, where one obtains a contribution from the Glasma graph, the mini-jet (BFKL) graph as well as a contribution from the interference between the two graphs. It has been shown that the latter contribution is zero in the dilute-dense framework~\cite{Dusling:2014oha}.  Following this guidance, we will compute the two contributions to Eq.\ref{eq_twocomp}. This issue is further discussed in Sec. 4.3.1 of the recent review article \cite{Dusling:2015gta}.

The quantity of experimental interest is the two particle correlation of charged hadrons that is given by
\begin{align}
\label{eq:dihadron}
&\frac{d^2N}{d\Delta \phi} = K \times \int\limits_{\eta_{\rm min}}^{\eta_{\rm max}} \!d\eta_p  \,d\eta_q \,\, {\cal A}\left(\eta_p,\eta_q\right) \\
&\!\!\times\int\limits_{p_T^{\rm min}}^{p_T^{\rm max}} \frac{dp_T^2}{2} \int\limits_{q_T^{\rm min}}^{q_T^{\rm max}}\frac{ d q_T^2}{2}\;\int d\phi_p \int d\phi_q\; \delta\left(\phi_p-\phi_q-\Delta\phi\right) \nonumber\\
&\!\!\times \int\limits_{z_0}^1\!\! dz_1 dz_2 \frac{D(z_1)}{z_1^2}\, \frac{D(z_2)}{z_2^2}
 \frac{d^2N_{}^{\rm \sl corr.}}{d^2\pp d^2\qp d\eta_p d\eta_q}\left(\frac{p_{\textrm{T}}}{z_1},\frac{q_{\textrm{T}}}{z_2},\Delta\phi \right)\nonumber
\end{align}
where $\eta_{\rm min, max}$ denote the pseudo-rapidity range of the detector and $\pp^{\rm min, max}, \qp^{\rm min, max}$ denote the transverse momentum ranges of trigger and associated particles. The function ${\cal A}\left(\eta_p,\eta_q\right)$ takes into account the acceptance of the uncorrelated background for direct comparison to experimental measurements. The exact form of ${\cal A}\left(\eta_p,\eta_q\right)$ for the CMS, ATLAS and ALICE collaborations has been studied in detail and can be found in Ref~\cite{Dusling:2013qoz}. As in the previous analyses the fragmentation function $D(z)$ is taken to be the NLO KKP~\cite{Kniehl:2000fe} parametrization~\footnote{There are two glasma graphs that are delta functions at $\Delta\Phi=0$ and $\Delta\Phi=\pi$.  These are given a natural width by finite source size corrections and final state broadening due to hadronization, corrections which are nonperturbative in nature and have been modeled in our fragmentation routines by replacing the angular delta function with a Gaussian.  Previous works had a typo for the form of the Guassian distribution used.  The correct function is 
\begin{equation} \delta(\phi_p-\phi_q)\to \frac{1}{\sqrt{\pi}\sigma} e^{-\frac{(\phi_p-\phi_q)^2}{\sigma^2}}
\end{equation}
with $\sigma=16\textrm{ GeV}^2/\pp^2$.  A similar expression is used for the delta function at $\phi_p-\phi_q=\pi$.  This typo did not affect any of the resulting calculations or conclusions. Regardless of the typo, we should stress that the functional form used for the smearing function does not affect the associated yield.}. 
The $K$-factor in Eq.~\ref{eq:dihadron} was taken as $K=1.5$ as in our previous analysis of p+p collisions at $\sqrt{s}=7$~TeV in order to accommodate higher order corrections, background uncertainties and sensitivity to the choice of fragmentation functions.

The number of reconstructed charged hadron tracks defined as 
\begin{align}
\Nch=\kappa_g \int\limits_{\eta_{\rm min}}^{\eta_{\rm max}} d\eta \int\limits_{\ptmin} d^2\pp \frac{dN}{d\eta \,d^2 \pp}\left(\pp\right)\,,
\label{eq:nch}
\end{align}
is estimated from the single inclusive gluon distribution
\begin{align}
\frac{\ud N}{\ud y_p\ud^2\pp }
&=\frac{8\alpha_s }{(2\pi)^6 \cf}\frac{S_\perp}{\pp^2}
\int\limits_{\kp} \Phi_{A}(\kp)\,\Phi_{B}(\pp-\kp)\,.
\label{eq:single}
\end{align}
In the above expression $\pp^{\rm min}$ is the minimum range of momentum for charged particle reconstruction in the experimental measurement. The ATLAS and CMS collaborations use $\pp^{\rm min}=0.4$ GeV, which indicates that $\Nch$ is dominated by soft processes whose uncertainties~\footnote{We  do not use any mechanism of fragmentation for the estimation of $\Nch$. Further, the infrared behavior of Eq.~\ref{eq:nch} is logarithmically sensitive to a mass term of order $\Lambda_{\rm QCD}$. In experiments, $\Nch$ is not corrected for detector efficiencies, so $\kappa_q \times S_\perp$  absorbs such uncertainties as well.} are absorbed into the nonperturbative constant $\kappa_g$ that represents a gluon to hadron conversion factor.  Because Eq.~\ref{eq:single} is proportional to $S_\perp$, the overall nonperturbative prefactor of Eq.~\ref{eq:nch} is $\kappa_g \times S_\perp$ and is fixed to reproduce the experimental value of $\Nch$ for the min-bias $\qnot^2=0.168$ GeV$^2$ multiplicity.

\begin{figure*}[htb]
\includegraphics[width=0.45\textwidth]{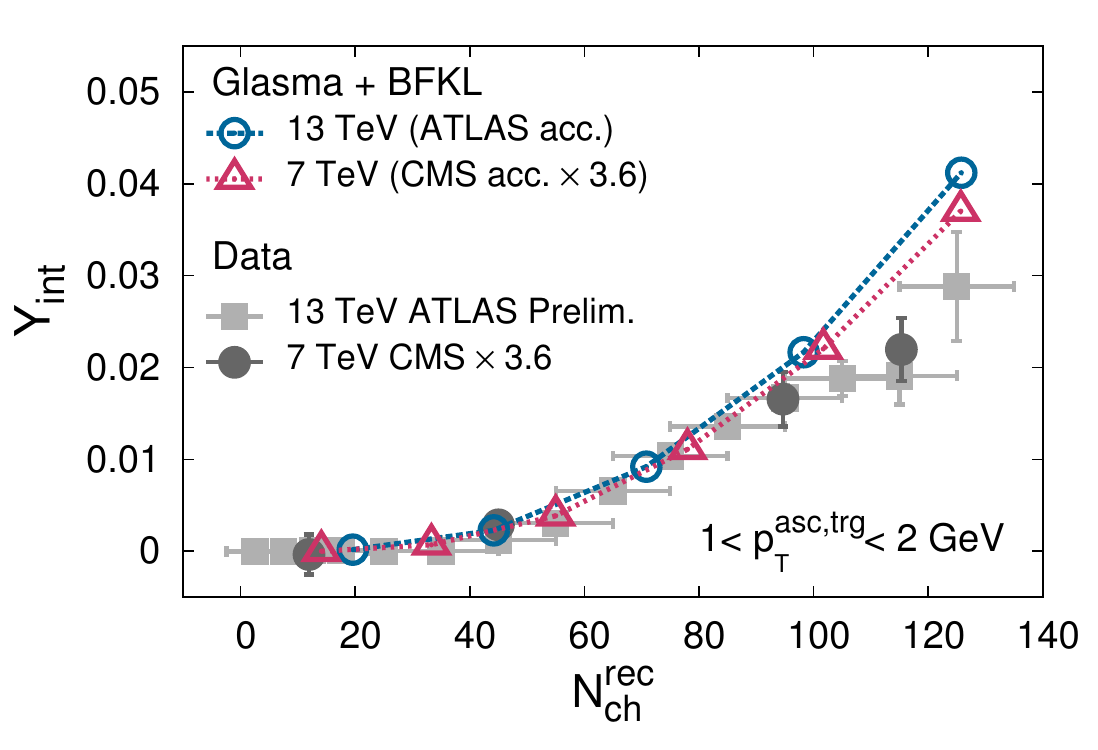}
\hspace{10pt}
\includegraphics[width=0.45\textwidth]{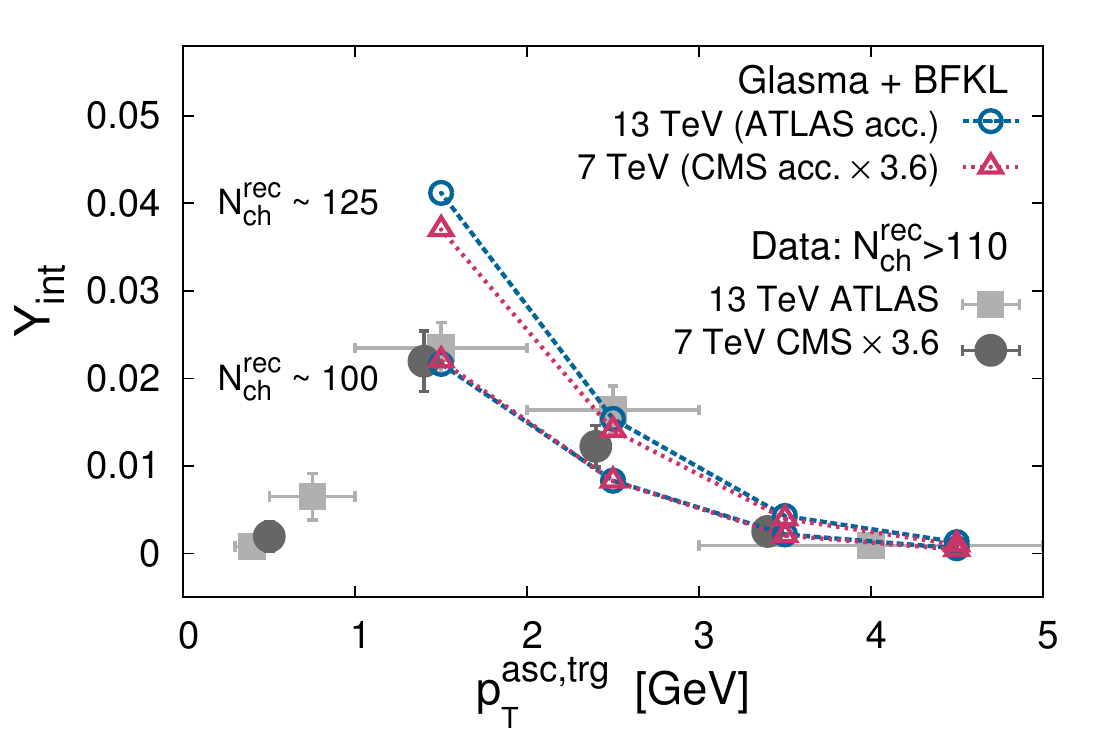}
\caption{Left: The near side integrated yield $Y_{\rm int}$ versus multiplicity at $\sqrt{s}=7$ and $13$ TeV.  Right:  $Y_{\rm int}$ versus $\pp^{\rm asc, trg}$ at $\sqrt{s}=7$ and $13$ TeV. The experimental data is for the multiplicity bin $\Nch>110$~\cite{1384640, Khachatryan:2010gv, CMS:2012qk}.  The lower theory curves have $\Nch\sim 100$ corresponding to $\qnot^2=0.840$~GeV$^2$ at 7~TeV and  $\qnot^2=0.672$~GeV$^2$ at 13~TeV.  The upper theory curves have $\Nch\sim 125$ corresponding to $\qnot^2=1.008$~GeV$^2$ at 7~TeV and  $\qnot^2=0.840$~GeV$^2$ at 13~TeV.}
\label{fig:yield} 
\end{figure*}
The variation of $\Nch$ only comes from the choice of $\qnot^2$ in our calculation, and as noted earlier, high multiplicity events can be modeled by using larger values of $\qnot^2$.  
We find the highest multiplicity bins analysed by CMS and ATLAS collaborations correspond to values of $\qnot^2$ that are 5-6 times its min-bias value.  This is consistent with the large values of $\qnot^2$ needed to explain the broad tail of the of the multiplicity distributions in~\cite{McLerran:2015qxa}; indeed as shown there, large fluctuations of the proton saturation scales are essential for a quantitative description of high multiplicity events in p+p collisions.  In the case of p+Pb collisions, the scenario is more complicated since there are two saturation scales involved in the problem~\cite{McLerran:2015lta}. 

The strength of the di-hadron correlation defined in Eq.~\ref{eq:dihadron} is normalized by the number of trigger particles defined as
\begin{align}
\Ntrig=\int\limits_{\eta_{\rm min}}^{\eta_{\rm max}} d\eta \int\limits_{\pp^{\rm min}}^{\pp^{\rm max}} d^2\pp\!\!\int\limits_{z_0}^1  dz \frac{D(z)}{z^2} \frac{dN}{d\eta \,d^2 \pp}\left(\frac{p_{\textrm{T}}}{z}\right)\, .
\label{eq:ntrig}
\end{align} 
In this study, we restrict our comparison with experimental data to $\pp>1$
GeV.  For $\pp\lesssim\Qs$ we expect corrections from multiple scattering to be important and
therefore do not show any results below 1~GeV. The two particle correlation has been computed only recently for all pT's using numerical classical Yang-Mills simulations~\cite{Schenke:2015aqa}. However, 
these results, which are obtained for gluon correlations, cannot be compared directly to data in the absence of a satisfactory implementation of a hadronization scheme. 

The free parameter $S_\perp$ in Eq.~\ref{eq:ntrig}  does not appear in the quantity of experimental interest--the integrated  associated yield per trigger, defined to be 
\begin{equation}
\label{eq:zyam}
\textrm{Y}_{\rm int} = \frac{1}{\Ntrig}\int\limits_0^{\Delta\phi_{\rm min.}} \!\!\!\!
d\Delta\phi\left(\frac{d^2N}{d\Delta\phi}-\left.\frac{d^2N}{d\Delta\phi}\right|_{\Delta\phi_{\rm
min}}\right)\, ,
\end{equation}
where $\Delta\phi=\Delta\phi_{\rm min.}$ corresponds to the minimum of the di-hadron correlation and the second term in the parenthesis of Eq.~\ref{eq:zyam} appears due to the ZYAM (zero yield at minimum) subtraction employed by the experiments.

With the above mentioned framework, and a few fully constrained parameters,
remarkable agreement was achieved for a wide range of p+p 7 TeV data without
the need for fine tuning. We will now extend our study of two particle
correlations to the recent ATLAS $\sqrt{s}$= 13 TeV data.  The dependence on
the center of mass energy \footnote{Since higher $\sqrt{s}$
corresponds to smaller $x$, for a fixed initial value of $\qnot^2$, the rcBK
equation leads to a higher saturation scale $\qs^2$ at larger $\sqrt{s}$;
$\qs^2(\qnot^2, \sqrt{s}\!=\!13~\text{TeV}) >\qs^2(\qnot^2,
\sqrt{s}\!=\!7~\text{TeV})$.} enters through the parton momentum fraction $x$
appearing in Eqs.~\ref{eq:glasma}, \ref{eq:bfkl} and \ref{eq:single}.  
In the CGC framework, it is $\qs^2$ that determines both single and double inclusive production in p+p collisions. Thus in both $\Nch$ and $Y_{\rm int}$ the dependence on the center of mass energy enters only through $\qs^2$, 
\begin{align}
\Nch (\qnot^2, \sqrt{s}) &= \Nch (\qs^2)\,,\\
Y_{\rm int} (\qnot^2, \sqrt{s}) &= Y_{\rm int} (\qs^2)\,,
\end{align}
where $\qs^2 \equiv \qs^2(\qnot^2, \sqrt{s})$. Fixing the value of $\Nch(\qs^2)$ fixes $\qs^2$ and therefore the value of $Y_{\rm int}(\qs^2)$. This leads to a universal scaling curve when $Y_{\rm int}$ is plotted against $\Nch$ at different energies as demonstrated in the left plot of Fig.~\ref{fig:yield}.
\begin{figure*}[th]
\includegraphics[width=0.45\textwidth]{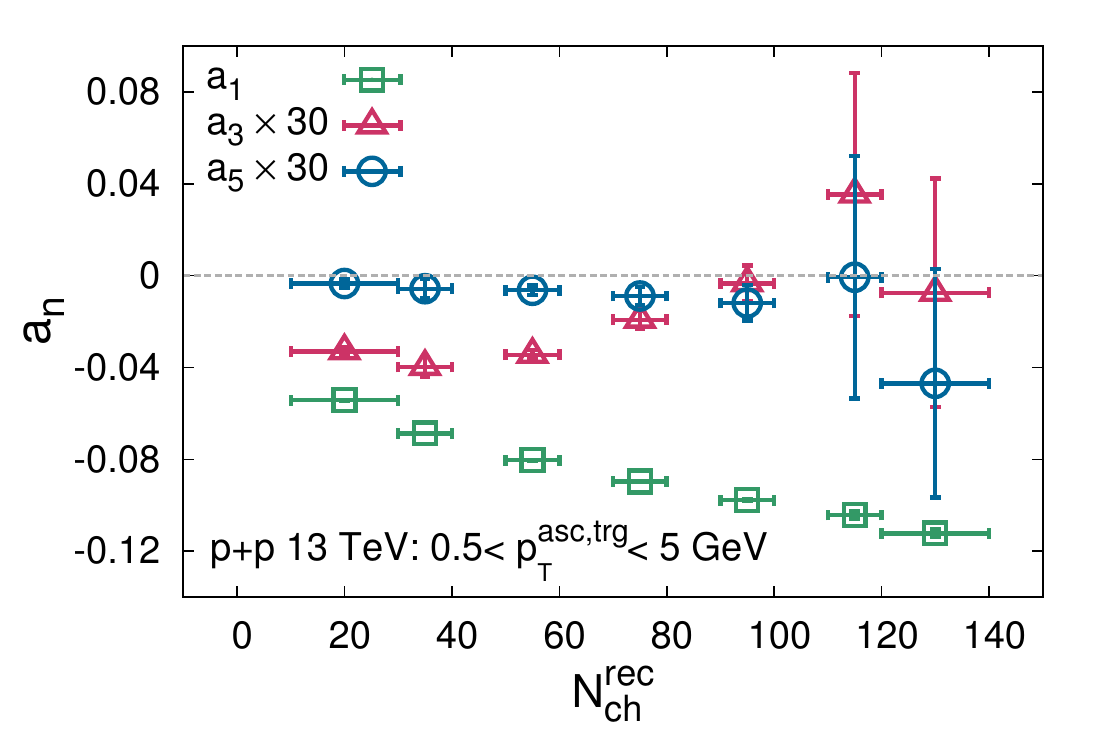}
\hspace{10pt}
\includegraphics[width=0.45\textwidth]{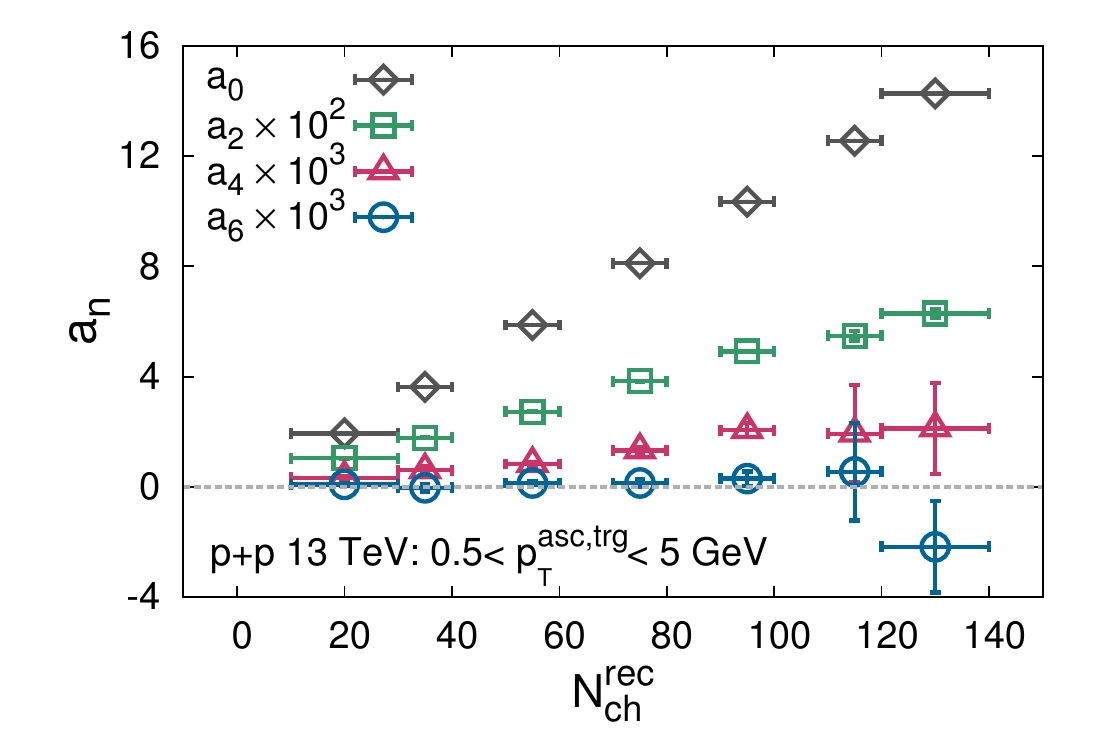}
\caption{The variation of odd (left) and even (right) harmonic coefficients $a_n$ defined in Eq.\ref{eq:an} extracted from the $\Delta\phi$ distribution of di-hadron yield $Y(\Delta \phi)$ from ATLAS p+p data~\cite{1384640} at $\sqrt{s}=$ 13 TeV over 2$<|\Delta\eta|<$5. The window of trigger and associated transverse momentum $0.5\!<\! \pp^{\rm asc, trg} \!<\! 5$ GeV. The $a_n$s are plotted against the reconstructed charged tracks $\Nch$. The point corresponding to $\Nch=130$ is measured for the multiplicity bin $\Nch>120$. (See text for details.)} 
\label{fig:an} 
\end{figure*}
On the same plot we show comparisons to recent preliminary data form the ATLAS collaboration at 13 TeV~\cite{1384640} along with the previously measured CMS data at 7 TeV~\cite{Khachatryan:2010gv, CMS:2012qk}. Our framework reproduces very well this scaling feature of the data. We hope similar measurements in p+p collisions at RHIC will enable one to test this prediction at lower energies albeit, it must be noted, the larger mean $x$ values at the lower energies may be sensitive to physics outside the regime of validity of this framework. 

The experimental data points shown in Fig.~\ref{fig:yield} for $\sqrt{s}$=13 TeV are obtained for an acceptance of $2\!<\!|\Delta\eta|\!<\!5$ and $-\Delta\phi_{\rm min}\!<\!\Delta\phi\!<\!\Delta\phi_{\rm min}$  whereas the results for $\sqrt{s}$=7 TeV is obtained for $2\!<\!|\Delta\eta|\!<\!4$ and $0\!<\!\Delta\phi\!<\!\Delta\phi_{\rm min}$~\cite{1384640}. The range of transverse momentum for associated and trigger particles is same for both energies and chosen to be $1<\pp^{\rm asc, trg}<2$ GeV. In order to take into account the acceptance difference in $\Delta\eta$ and $\Delta\phi$ the values of $Y_{\rm int}$ at $\sqrt{s}$=7 TeV are scaled by a constant factor of 3.6. We follow the same approach for the calculation of $Y_{\rm int}$ in our framework for direct comparison with the data.  

We match the value of the $\Nch$ at $\sqrt{s}=$ 7 TeV  in our computation for $\qnot^2=0.168$ GeV$^2$  to the min-bias value of CMS data which is $\Nch=14$. We use this normalization to obtain higher $\Nch$ by increasing the value of $\qnot^2$ in integer multiples of $0.168$ GeV$^2$ for both energies studied here. Over the range of $\Nch$ shown in Fig.~\ref{fig:yield}, the six points in our calculation for $\sqrt{s}=7$ TeV correspond to values of $\qnot^2$ in the range of $0.168\!-\!1.008$ GeV$^2$ and the five points for $\sqrt{s}=13$ TeV correspond to a variation of $\qnot^2$ in the range of $0.168\!-\!0.840$ GeV$^2$. The measurements of $\Nch$ for both CMS and ATLAS do not include efficiency corrections. Since we use the same normalization factor for comparison to both CMS and ATLAS data, the relative difference in the efficiencies for these two experiments gives rise to about a  $5\%$ uncertainty in the normalization of our $\Nch$ with the experimental data.  Furthermore, it is not clear that the min. bias configuration of the proton obtained from HERA DIS data is the same as the min. bias configuration of the proton in hadronic collisions.  There is thefore a considerable systematic uncertainty (which we estimate~\footnote{Another issue relevant to the scaling shown in Fig.~\ref{fig:yield} is as follows. The $\Nch$ measurement by the ATLAS collaboration uses a coverage of  $-2.5\!<\!\eta\!<\!2.5$ whereas CMS collaboration uses $-2.4\!<\!\eta<2.4$, this difference in acceptance can also lead to an approximately $5\%$ difference in the min-bias value of $\Nch$ at a given energy. This corresponds to a systematic uncertainty in the comparison of the results at two energies which is common for both data and our calculation. This small uncertainty will be resolved when the data from the CMS collaboration is available and a plot similar to Fig.\ref{fig:yield} is obtained using the same acceptance ($-2.4\!<\!\eta<2.4$) for two different energies.} to be on the order of $10\%$) of the scaling of the x-axis in the left plot of Fig.~\ref{fig:yield}. We avoid such fine tuning since it does not affect the qualitative features of the results presented here. 

In the right plot of Fig.~\ref{fig:yield}, we show the variation of $Y_{\rm int}$ with the transverse momentum of trigger and associated particles for a fixed bin of $\Nch >110$. This bin of multiplicity corresponds to $0.672 \le \qnot^2 \le 0.840$ GeV$^2$ for $\sqrt{s}=$ 13 TeV and $0.840 \le \qnot^2 \le 1.008$ GeV$^2$ for $\sqrt{s}=$ 7 TeV in our calculation. The values of $Y_{\rm int}$ corresponding to these two values of $\qnot^2$ give rise to the two sets of curves shown here. Within the regime of validity of our calculation ($\pp^{\rm asc, trg} > 1$ GeV),  the agreement with the data is very good. The two sets of results shown in Fig.~\ref{fig:yield} indicate that for p+p collisions $Y_{\rm int}$ only depends on $\Nch$. As noted previously, a natural explanation is that, for symmetric systems, multiparticle production is determined by a single scale $\qs^2$. 

In asymmetric systems like p+Pb two different saturation scales control multiparticle production. Since the evolution of the proton scale with energy is different from that of nuclei, a complicated relation between the two saturation scales will determine the correlation of $\Nch$ and $Y_{\rm int}$. It will therefore be very interesting to see the same measurement of $\Nch$ versus $Y_{\rm int}$ in p+Pb collisions at two different center of mass energies. This may be feasible with forthcoming data from 200 GeV p+A collisions at the RHIC and in the future at the highest p+A energies of the LHC.  A detailed calculation in our framework is left for future work.  

We will consider now another important feature of the p+p data at $\sqrt{s}=$13 TeV measured by the ATLAS Collaboration.  In Ref~\cite{1384640}, the $\Delta\phi$ dependence of $Y(\Delta\phi)$ is provided in different bins of multiplicity for $0.5<\pp^{\rm asc, trg}<5$ GeV.  The full $\Delta\phi$ distribution contains a lot more information than the integrated $Y_{\rm int}$.
We performed a decomposition of the azimuthal structure of the two particle correlation by fitting the experimental data to the functional form
\begin{equation}
Y(\Delta\phi) = a_0 + \sum\limits_{n=1}^{10} 2 a_{n} \cos{n \Delta \phi}.
\label{eq:an}
\end{equation}
In Fig.~\ref{fig:an} we show the results of fitting the ATLAS data by plotting the different $a_n$s as a function of $\Nch$. 
The Fourier components of the two particle correlation show a remarkable pattern; $a_n \apprle 0$ for the odd harmonics and $a_n \apprge 0$ for the even harmonics up to the highest multiplicity window of $\Nch > 120$.  The $a_0$ contribution is the uncollimated background per trigger and scales linearly with $\Nch$--this is natural since this corresponds to the uncollimated double inclusive yield scaling as $(\Nch)^2$.  The negative value of $a_1$ is also to be anticipated since it corresponds to the azimuthal correlation of mini-jets. The $a_2$ contribution is the ridge and is clearly positive--in a hydrodynamic scenario, its square root would approximately correspond to the root-mean-square elliptic flow coefficient $v_2$.  Most interestingly, the $a_3$ contribution is negative.  The consequences of this can be understood by the relationship between the $a_n$'s and the $v_n$'s as obtained from the single particle distributions with respect to a reaction plane,
\begin{align}
\frac{a_n(\pp,y_p,\qp,y_q)}{a_0(\pp,y_p,\qp,y_q)}=v_n(\pp,y_p)v_n(\qp,y_q)\;.
\end{align}
From the above, it is very difficult to acquire a negative $a_3$ in hydrodynamics as it is a product of $v_3$'s at different rapidities.  For a boost invariant expansion, $a_3$ is necessarily positive definite.  Indeed, this factorization is observed in Pb+Pb collisions and provides strong evidence for collective behavior in nucleus-nucleus (A-A) collisions.  The even-odd pattern continues for $a_{4,5,6}$--these are clearly very small but appear distinct within the impressive statistical accuracy of the ATLAS measurement. 

A two particle correlation measurement in hydrodynamics (which requires event-by-event simulations) would require all $a_n$'s to be positive definite. The ATLAS measurement may therefore appear to conclusively rule out a hydrodynamic interpretation of the high multiplicity p+p data. However if one construes $a_3$ as resulting from a ``mini-jet" mechanism plus a ``hydro medium" mechanism,  it is conceivable that a relatively large negative $a_3$ from the mini-jets might mask a positive $a_3$ from the hydro mechanism which would grow with $\Nch$ if the mini-jet contribution is nearly constant. But such a distinction between mini-jet and medium is completely artificial and requires fine tuning. Particles of a given $p_T$ do not carry labels as belonging to the one or the other particle production mechanism and rescatterings should be expected to affect both equally. 

A consistent treatment of both types of two particle correlations is available within the CGC/glasma framework itself, which also provides the mechanism, in large systems, for matter to thermalize and flow hydrodynamically. The glasma graph approximation to the full formalism is a good approximation when $\Nch$ is not too large. In this approximation, the anticipated ordering of the $a_n$ moments is precisely along the lines of what one sees in Fig.~\ref{fig:an}. 

However when $\Nch$ becomes large, coherent multiple scattering effects within domains of size $1/Q_S$  should become important, requiring numerical solutions of Yang-Mills equations that account for such effects. In \cite{Schenke:2015aqa}, it was demonstrated that they can lead to a positive $a_3$. It has not yet been shown that this $a_3$ dominates the one from mini-jets computed in the same framework. The appearance of a positive $a_3$ in p+A collisions, in contrast to p+p collisions at the same $\Nch$, may indicate plausibly that this coherent multiple scattering mechanism is precocious in p+A collisions. 

Despite the many caveats noted here, and earlier in the text, we believe it is fair to say the high multiplicity p+p results challenge the hydrodynamic interpretation of the p+p ridge. In contrast, the data are consistent qualitatively and even quantitatively with the expectations of the glasma initial state framework. Further clarity can be achieved once high multiplicity data from RHIC becomes available. A finer binning in $p_T$ of the $a_n$ moments, the extraction of the two particle yields to higher values of $\Nch$ and the possible extraction of 4-particle cumulants would also be helpful. While these are experimentally difficult to extract in p+p collisions, they would add clarity to the ongoing debate on the nature of the strongly correlated QCD matter produced in small systems. 
 
\section*{Acknowledgement}
We thank Soumya Mohapatra and  Dennis Perepelitsa of the ATLAS collaboration for useful information regarding the ATLAS data. We thank Bj\"{o}rn Schenke, S\"{o}ren Schlichting, Paul Sorensen and Takahito Todoroki for discussions. R. V. thanks Jean-Yves Ollitrault for helpful correspondence. The authors are supported under Department of Energy contract number Contract No. DE-SC0012704. R. V. would like to thank the Institut f\"{u}r Theoretische Physik, Heidelberg for their kind hospitality and the Excellence Initiative of Heidelberg University for their support. This research used resources of the National Energy Research Scientific Computing Center, a DOE Office of Science User Facility supported by the Office of Science of the U.S. Department of Energy under Contract No. DE-AC02-05CH11231.
 
\bibliographystyle{apsrev4-1}
\bibliography{ppridge_v2}

\end{document}